\documentclass{appolb}
\usepackage{epsfig}

\newcommand{\del}{\mathrm{d}}
\newcommand{\figurewidth}{11cm}
\newcommand{\smallfigurewidth}{6.8cm}
\newcommand{\halffigurewidth}{6cm}

\begin{document}
\pagestyle{plain}

\eqsec
\newcount\eLiNe\eLiNe=\inputlineno\advance\eLiNe by -1

\title{PRODUCTION OF K$^+$-MESONS\\IN PROTON-NUCLEUS INTERACTIONS
}
\author{Markus \underline{B\"USCHER}\thanks{\tt m.buescher@fz-juelich.de}
         and Mikhail NEKIPELOV\thanks{also working at Petersburg Nuclear Physics Institute, 
                                      188350 Gatchina, Russia}\\
         for the ANKE collaboration
\address{Institut f\"ur Kernphysik, Forschungszentrum J\"ulich,
52425~J\"ulich, Germany}}
\maketitle

\begin{abstract}
  The production of $K^+$-mesons in $pA$ ($A=$ C, Cu, Ag, Au)
  collisions has been investigated at the COoler SYnchrotron
  COSY-J\"ulich at beam energies $T_p=1.0 \ldots 2.3$ GeV using the
  ANKE spectrometer. ANKE allows to measure double differential cross
  sections for $K^+$-production at forward angles $\vartheta<
  12^{\circ}$ in a wide momentum range.  The cross sections obtained
  in these experiments together with those from literature show a
  uniform behavior as functions of momentum transfer and excitation
  energy of the residual nucleus. At the lowest measured beam energy,
  far below the free $NN$-threshold ($T_{NN}=1.58$ GeV), the data
  reveal a high degree of collectivity in the target nucleus. The
  spectra obtained above threshold should allow one to extract the
  nuclear $K^+$-potential with an accuracy of better than $\pm3$~MeV.
  Information about the mechanisms leading to $K^+$-production can be
  obtained from $K^+p$ and $K^+d$ coincidence measurements.
\end{abstract}

\section{Inclusive $K^+$-momentum spectra from ANKE}
\label{sec:inclusive}
A central topic of hadron physics is to study the influence of the
nuclear medium on elementary processes, for example by measuring meson
production at projectile energies below the threshold for free $NN$
collisions (so-called subthreshold production). These processes
necessarily involve collective effects of the nucleons inside the
target nucleus.  $K^+$-production is particularly well suited for such
investigations since the meson is rather heavy, as compared to pions,
so that its production requires strong medium effects.  Final state
interactions of $K^+$ mesons in nuclei are generally considered to be
rather small, due to their strangeness of $S=+1$.  As a consequence,
the production of $K^+$-mesons in proton-nucleus collisions is of
great importance to learn about either cooperative nuclear phenomena
or high momentum components in the nuclear many-body wave function.

The COoler SYnchrotron COSY-J\"ulich~\cite{cosy}, which provides
proton beams in the range $T_p = 0.04 \ldots 2.83$~GeV, is well suited
for the study of $K^+$-production. In measurements with very thin
windowless internal targets, secondary processes of the produced
mesons can be neglected and, simultaneously, sufficiently high
luminosities of more than $L=10^{32}\ \mathrm{cm^{-2}s^{-1}}$ can be
obtained.  Subthreshold $K^+$-production was a prime motivation for
building the large-acceptance ANKE spectrometer~\cite{ANKE_NIM} within
one straight section of the COSY ring.  It consists of three dipole
magnets, which separate forward-emitted reaction products from the
circulating proton beam and allow to determine their emission angles
and momenta.  Depending on the choice of the magnetic field in the
spectrometer dipole, $K^+$-mesons in the momentum range $p_K\approx
150\ldots600$ MeV/c can be detected.  The layout of the device,
including detectors and the data-acquisition system, was optimized
\cite{K_NIM} to study $K^+$-spectra down to $T_p \approx
1.0$~GeV. This is a very demanding task because of the small
$K^+$-production cross sections, e.g.\ 39~nb for $p$C collisions at
1.0~GeV~\cite{pnpi}.  Thus, the kaons have to be identified
\cite{K_NIM} in a background of secondary protons and pions which is
up to $10^6$ times more intense.

\begin{figure}[ht]
  \begin{center}
    \vspace*{-3mm}
    \resizebox{\smallfigurewidth}{!}{\includegraphics[scale=1]{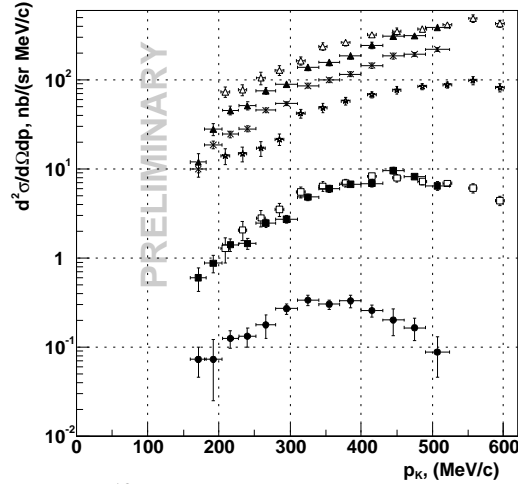}}
    \vspace*{-3mm}
    \caption{Double differential $p^{12}C\rightarrow K^+X$  cross sections 
      measured at ANKE.  Black circles denote the cross sections
      measured at $T_p=1.0$ GeV~\cite{ANKE_1.0GeV}, squares at
      $T_p=1.2$ GeV, stars at $T_p=1.5$ GeV, crosses at $T_p=2.0$ GeV
      and triangles at $T_p=2.3$ GeV. All closed symbols correspond to
      measurements with $B=1.3$~T and $\vartheta<12^{\circ}$,
      open symbols to $B=1.6$~T and $\vartheta_\mathrm{vert.}<
      3.5^{\circ},\,\,\vartheta_\mathrm{hor.} <6^{\circ}$.  The error
      bars are purely statistical.}
    \label{fig:cross_sections}
  \end{center}
\end{figure}

The measured cross sections for $K^+$ production in $p$C interactions
are shown in Fig.~\ref{fig:cross_sections}. The cross sections were
normalized to known pion-pro\-duction cross sections~\cite{papp,abaev}
measured for similar kinematical conditions~\cite{meson2000}. For
$T_p=1.2$ and 2.3~GeV two data sets have been obtained for different
settings of the field strength in the spectrometer dipole, $B=1.3$~T
and 1.6~T. These two modes of operation allow to explore slightly
different but overlapping regions of the kaon-momentum spectra.  At
$T_p=1.0$ GeV ($B=1.3$~T), ANKE for the first time allows to measure
the full momentum spectrum of the produced kaons at deep-subthreshold
energies~\cite{ANKE_1.0GeV}.  The geometry of the experimental setup
as well as the analysis procedures are identical for both field
values.  Fig.~\ref{fig:cross_sections} shows that there are
discrepancies between the two data sets for $p_K\leq400$~MeV/c at
$T_p=2.3$ GeV, whereas at $T_p=1.2$~GeV they coincide.  We conclude
that there is an as yet unknown source of systematic errors, which may
yield an additional uncertainty of up to 30\% at the higher beam
energies. It should be noted that these uncertainties vanish when
one calculates the cross-section {\em ratios} for different target
nuclei \cite{PLB}.  Thus, these ratios are used for the discussion of
reaction mechanisms and final-state interaction effects in
Sects.~\ref{sec:mechanisms} and \ref{sec:fsi}.

\section{Systematics of available data}
\label{sec:systematics}
Table~\ref{tab:syst} shows the available data on $K^+$-production in
$pA$ collisions. The different data sets were obtained for
non-overlapping kinematical parameters (i.e.\ beam energies, kaon
emission angles and momenta) which prevents a direct comparison with
the data from ANKE and a test of the overall normalizations.

\begin{table}[h]
  \caption{Data on $K^+$-production in $pA$ collisions (ordered by the year of
           publication) at various beam energies $T_p$, kaon momenta $p_K$ and 
           emission angles $\vartheta_K$.}
    \label{tab:syst}
    \begin{center}
    \begin{tabular}{c|c|c|c|c}
      $T_p$ (GeV) & Targets & $p_K$ (MeV/c) & $\vartheta_K$ ($^{\circ}$)& Measured at\rule[-2mm]{0mm}{2mm}\\
      \hline
      \rule[2mm]{0mm}{2mm}
      0.842\ldots0.99 & Be\ldots Pb&\multicolumn{2}{c|}{\em total cross sections} & PNPI \protect\cite{pnpi} \\
      2.1 & NaF, Pb&350--750 & 15--80 & LBL \protect \cite{schnetzer} \\
      1.2, 1.5, 2.5 & C, Pb&500--700 & 40   & SATURNE \protect\cite{debowski} \\
      1.2 & C &165--255 & 90   & CELSIUS \protect\cite{badala} \\
      1.7\ldots2.91& Be& 1280 & 10.5  & ITEP \protect \cite{akindinov} \\
      2.9 & Be&545 & 17  & ITEP \protect \cite{buescher} \\
      1.0 & C, Cu, Au&171--507 & $\leq12$  & {\bf ANKE} \protect\cite{ANKE_1.0GeV}\\
      2.5, 3.5 & C, Au& 300--1050   & 40  & KAOS, \cite{scheinast}\\
      1.2, 1.5, 2.0, 2.3 &C,Cu,Ag,Au\hspace*{-1mm}&171--595 & $\leq12$  & {\bf ANKE}, {\em prelim.}
    \end{tabular}
    \vspace*{-2mm}
    \end{center}
\end{table}

\begin{figure}[htbp]
  \begin{center}
    \vspace*{-2mm}
    \resizebox{\halffigurewidth}{!}{\includegraphics[scale=1]{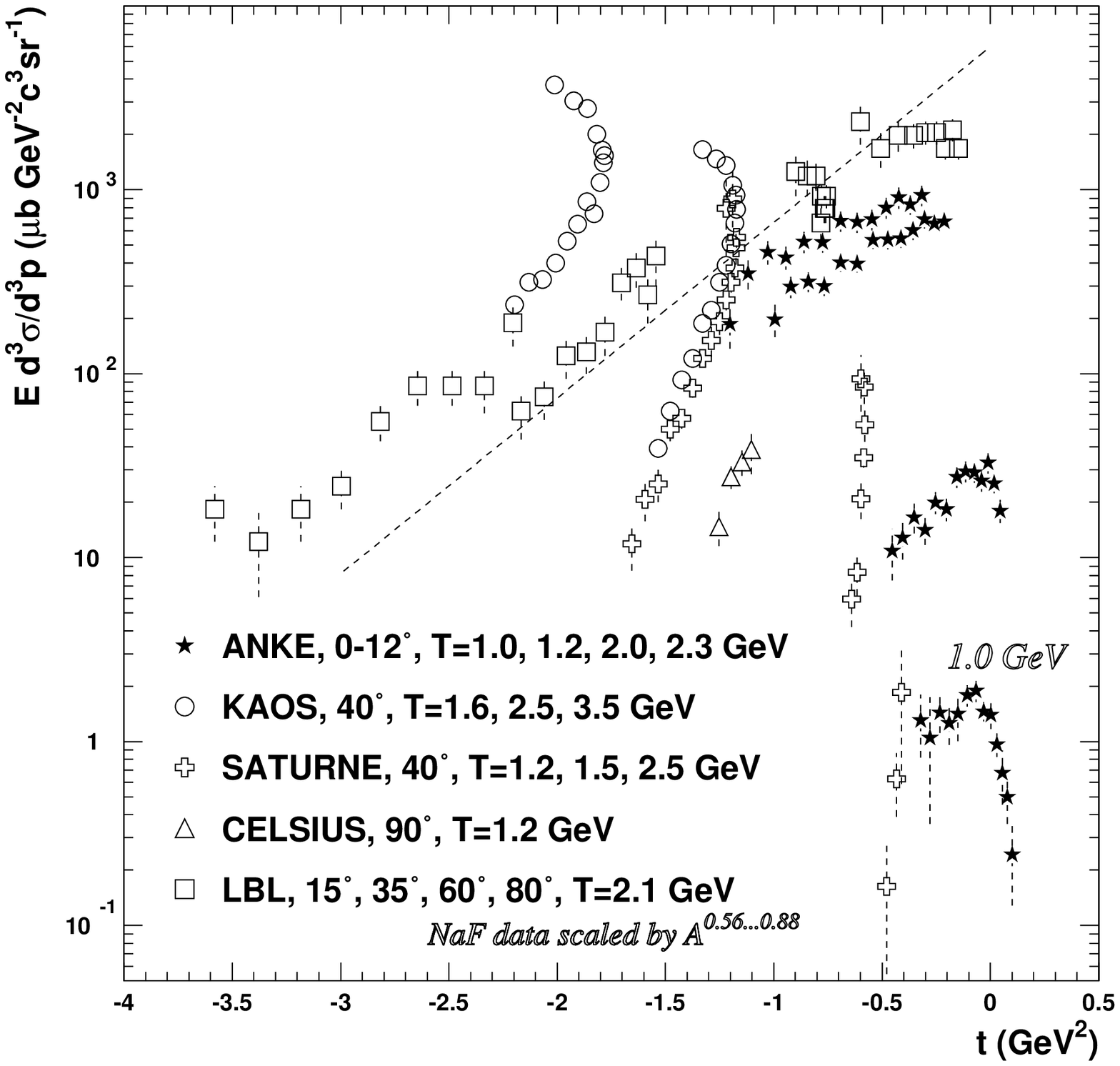}}
    \resizebox{\halffigurewidth}{!}{\includegraphics[scale=1]{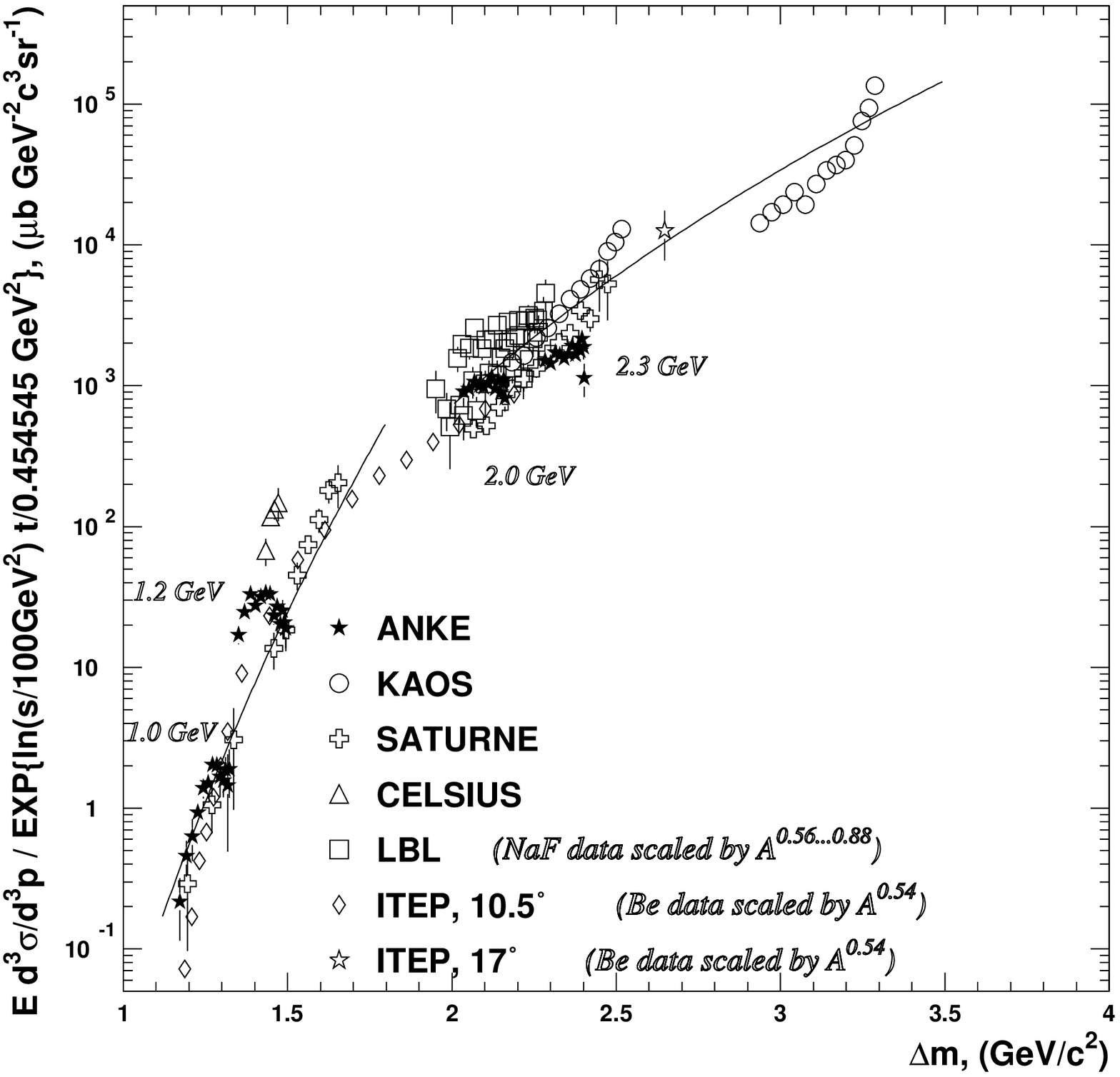}}
    \caption{Left figure: Invariant $p^{12}C\rightarrow K^+X$ cross
      sections as a function of the four-momentum transfer $t$ between
      the beam proton and the outgoing kaon. Right figure: Scaled
      invariant $p^{12}C\rightarrow K^+X$ cross sections as a function
      of the excitation energy $\Delta m$ of the target nucleus. The
      data from LBL \protect{\cite{schnetzer}} and ITEP
      \protect{\cite{akindinov,buescher}} were measured with $NaF$
      ($Be$) targets and have been corrected for the target-mass dependence
      according to Ref.\protect{\cite{syst}}. The beam energies of the
      ANKE data sets (filled stars) are indicated; for $T_p=2.3$ GeV
      the data obtained at $B=1.3$~T (closed triangles in
      Fig.~\ref{fig:cross_sections}) were used.}
    \label{fig:systematics}
  \end{center}
\end{figure}

In a recent publication \cite{syst}  we showed that the measured
invariant cross sections $E\,\del^3\sigma/\del^3p$ follow an
exponential scaling behavior when plotted as a function of the
four-momentum transfer $t$, which is illustrated in
Fig.\ref{fig:systematics} (l.h.s.).

Apart from the data taken with ANKE at $T_p=1.0$ and 1.2 GeV all
spectra cover the range of negative $t$.  The sharp fall-off of the
cross sections from ANKE towards positive values of $t$ was explained
\cite{syst} by the fact that the data were taken very close to the
kinematical limit for hyper-nucleus formation at $t=0.145\ 
\mathrm{GeV}^2$ ($T_p=1.0$ GeV) which is expected to be accompanied by
very small cross sections. For $t<0$ the different data follow an
exponential dependence of:
\begin{equation}
E\, \frac{\del^3\sigma}{\del^3p} = c_0\exp{[b_0t]}\ ,
  \label{eq:tdep1}
\end{equation}
with parameters $c_0$ and $b_0$ given in Ref.\cite{syst}. It has been
speculated \cite{syst} that deviations from the exponential behavior
(see, e.g., the data from KAOS \cite{scheinast} in
Fig.~\ref{fig:systematics}) reflect a dependence on the available
squared CM energy $s$ and the excitation energy of the target nucleus,
$\Delta m=m_X-m_A$ ($m_X$ and $m_A$ denote the mass of the target
nucleus before and after the reaction process, respectively). Based on
Regge phenomenology the following formula has been suggested as a
parameterization of the invariant cross section:
\begin{equation}
  E\, \frac{\del^3\sigma}{\del^3p} \propto f(t,m_X^2) \exp{[b_0t\cdot\ln{(s/s_0)}]}\ .
  \label{regge}
\end{equation}
The r.h.s.\ of Figure \ref{fig:systematics} shows the invariant cross
sections, divided by an average $s,t$ dependence, using $b_0=2.2\ 
\mathrm{GeV}^{-2}$ (indicated by the dashed line on the l.h.s.\ of
Fig.\ref{fig:systematics}) and $s_0=100\ \mathrm{GeV}^{2}$, as a
function of the excitation energy $\Delta m$.  The solid lines in
Fig.\ref{fig:systematics} correspond to a parameterization of the
invariant cross sections in $p^{12}$C interactions
\begin{equation}
  \label{eq:parametrization}
  E\, \frac{\del^3\sigma}{\del^3p}
      = \sigma_0\cdot \Delta m^{N_0} \cdot \exp{[b_0t\cdot\ln{(s/s_0)}]} 
      = \sigma_0\cdot \Delta m^{N_0} \cdot (s/s_0)^{b_0t} \ ,
\end{equation}
with the following fitted parameter sets:
\begin{table}[h]
  \caption{Parameters obtained from a fit with Eq.~(\ref{eq:parametrization})}
    \label{tab:para}
    \begin{center}
    \begin{tabular}{c||c|c|c|c}
     Beam energy & $\sigma_0$  & $N_0$~$(n)$ & $b_0$ & $s_0$ \\
     \rule[-2mm]{0mm}{2mm}(GeV) & ($\mathrm{nb}\ \mathrm{GeV}^{-2}\mathrm{c}^{3}\mathrm{sr}^{-1}$)
                                            & & (GeV$^{-2}$) &  (GeV$^2$)\\
      \hline
      \rule[3mm]{0mm}{3mm}
        $<1.58$ & 25 & 17 (13) &     &     \\
                                    &    &    & 2.2 &  100\\
        $>1.58$ &1000&9.5 (8)&     &     \\
    \end{tabular}
    \end{center}
\end{table}

The term $\Delta m^{N_0}$ reflects a phase space behavior $\sigma
\propto \Delta m^{N_0}=\Delta m^{(3n - 5)/2}$ with $n$ particles in
the final state. As can be seen from Fig.\ref{fig:systematics}, the
para\-meteri\-zation from Eq.~(\ref{eq:parametrization}) can describe
all available data on $K^+$-production in $pA$ reactions obtained in
largely different angular-momentum intervals within a factor $\sim2-3$.
This parameterization is based on Lorentz invariant variables and,
thus, is independent of the choice of the reference system, $pp$ or
$pA$.

\section{Kinematical considerations} 
\label{sec:kinematics}
The parameterization given by Eq.~(\ref{eq:parametrization}) not only
supplies a useful description of all available data, but
Fig.~\ref{fig:systematics} (r.h.s.) also reveals that
\begin{enumerate}
\item The data from ANKE at $T_p=1.0$ GeV were obtained down to
  $\Delta m=1.173$ GeV/c$^2$, i.e.\ very close to the kinematical
  limit at $\Delta m_\mathrm{min}=m_\Lambda=1.116$ GeV/c$^2$.  At this
  limit no energy can be transfered which excites the target nucleus
  or knocks out target nucleons. The target nucleus must take part in
  the reaction as a whole such that the effective target mass is
  $12\cdot m_N$. There are only two particles, $K^+$-meson and
  nucleus, in the final state ($n=2$).
\item Within a phase-space treatment of the final state, the steep
  rise ($N_0=17$, $n=13$) of the subthreshold data indicates that all
  12 nucleons plus the $\Lambda$-hyperon carry away energy in the
  final state. This seems to be in contradiction with the above
  statement of $n=2$.  However, this might reflect that more and more
  nucleons in the {\em initial state} must take part in the
  kaon-production process when approaching the kinematical limit (see
  discussion of the 1.0 GeV data below).
\item The different slope parameters $N_0$ below and above the free
  $NN$ threshold ($T_{NN}=1.58$ GeV) indicate a change of the dominant
  $K^+$-production mechanism. This has been predicted by various model
  calculations \cite{cassing,roc,sibirtsev,paryev} and is confirmed by
  recent coincidence measurements with ANKE (see
  Sect.~\ref{sec:mechanisms} for details).
\end{enumerate}

\begin{figure}[ht]
  \begin{center}
    \vspace*{-3mm}
    \resizebox{\figurewidth}{!}{\includegraphics[scale=1]{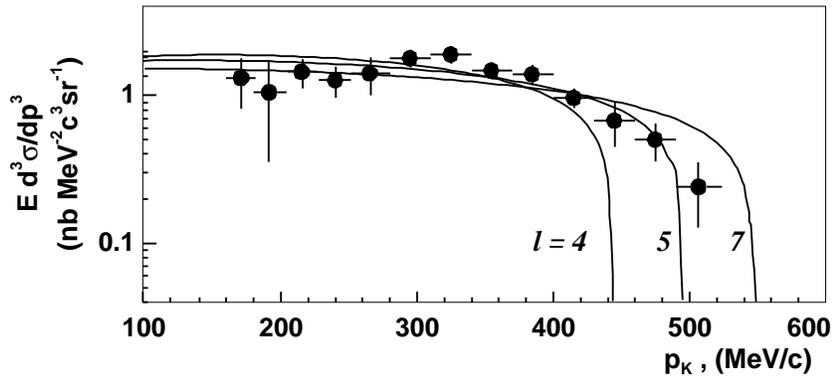}}
    \caption{Invariant $K^+$-production cross section 
      for $p(1.0\ \mathrm{GeV})^{12}C \rightarrow K^+X$ as a function
      of the $K^+$-momentum. The solid lines describe the behavior of
      the invariant cross section within a phase-space approximation
      (Eq.~(\ref{eq:kinem})). The figure has been taken from Ref.\ 
      \protect{\cite{ANKE_1.0GeV}}.}
  \label{fig:1.0GeV}
  \end{center}
\end{figure}

The ANKE data for $K^+$ production at $T_p=1.0$~GeV are plotted as
invariant cross sections in Fig.~\ref{fig:1.0GeV}, see also Ref.\ 
\cite{ANKE_1.0GeV}. As discussed above the high-momentum part of the
kaon spectrum (smallest values of $\Delta m$) is most sensitive to
collective nuclear effects. For example, if the kaons are produced in
a collision of the beam proton with a single nucleon, internal momenta
of at least $p_N \approx 550$~MeV/c are needed in order to produce
kaons in the forward direction with momenta of $p_K\approx 500$~MeV/c.
Such high momentum components above $p_N\approx 500$~MeV/c are
essentially due to many-body correlations in the nucleus.

To get a {\em rough estimate} of the number of participating nucleons,
one can describe the invariant cross section within a phase-space
approximation. The invariant cross section for the
$p{+}(lN){\to}(lN){+}\Lambda{+}K^+$ reaction is then
\begin{eqnarray}
  \label{eq:kinem}
  E\, \frac{\del^3\sigma}{\del^3p} &\propto&
  \frac{\sqrt{(s_l-m_\Lambda^2 - l^2m_N^2)^2-4m_\Lambda^2 l^2m_N^2}}{s_l}\nonumber\\
  &s_l&=s{+}m_K^2{-}2E_K(T_p{+}[l{+}1]m_N){+}2p_Kp_pcos\theta_K\ ,
\end{eqnarray}
where  $s$ denotes the square of the CM
energy of the incident proton and the $l$ target nucleons. The solid
lines in Fig.~\ref{fig:1.0GeV} show the invariant cross section
calculated with Eq.~(\ref{eq:kinem}) for $l=4,5,7$.  Although this
approach completely neglects the intrinsic motion of the $l$ target
nucleons, it shows that kaon production at $T_p=1.0$~GeV can only be
understood in terms of cooperative effects, where the effective number
of nucleons involved in the interaction is $\sim5-6$. It has been
suggested \cite{pnpi,cassing,roc,sibirtsev,paryev} that such effects
can be described in terms of multi-step mechanisms (see
Sect.~\ref{sec:mechanisms}) or high-momentum components in the nuclear
wave function.

\section{Production mechanisms}
\label{sec:mechanisms}
The dependence of meson-production cross sections on the target mass
$A$ has frequently been used (see Ref.~\cite{syst} and references
therein) to obtain information about production mechanisms. At large
beam energies, $K^+$-mesons are supposed to be dominantly produced in
a collision of the beam proton with a single target nucleon
($pN\rightarrow N K^+\Lambda$). In this case one expects that the
cross section scales as $A^{2/3}$.  Below threshold multi-step
production, in particular two-step reactions with the creation of
intermediate pions ($pN_1\rightarrow NN\pi$, $\pi N_2\rightarrow
K^+\Lambda $), should dominate, and the cross sections are then
proportional to $A^1$ \cite{syst}.

\begin{figure}[ht]
  \begin{center}
    \resizebox{\figurewidth}{!}{\includegraphics[scale=1]{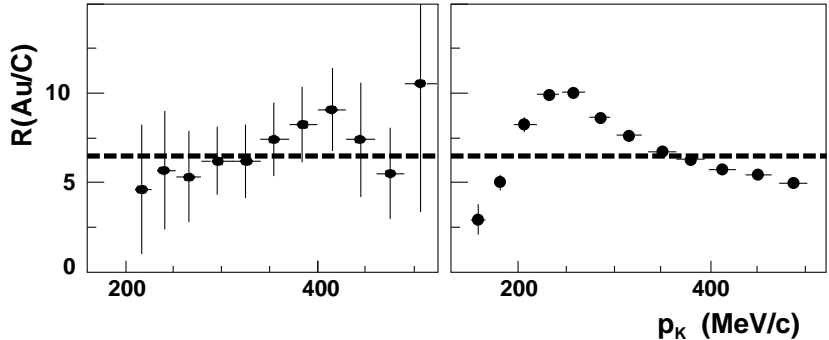}}
    \vspace*{-2mm}
    \caption{Ratios of the $K^+$ production cross sections Au/C measured at 
      $T_p=1.0$ GeV (l.h.s.) \cite{ANKE_1.0GeV} and 2.3 GeV (r.h.s.)
      \cite{PLB} as a function of the kaon momentum. The dashed line
      indicates an $A^{2/3}$ behavior of the cross sections.}
  \label{fig:ratios}
  \end{center}
\end{figure}

Figure \ref{fig:ratios} shows the cross section ratios for Au and C
nuclei at $T_p=1.0$ GeV and 2.3 GeV. Surprisingly, at both beam
energies the momentum averaged cross-section ratios seem to be in line
with $\del \sigma/\del\Omega(A)\propto A^{2/3}$. There is, at least,
no increase of the average ratio towards the lower beam energy.  The
mass dependence of the 1.0~GeV data from ANKE, selecting forward
angles $\vartheta<12^{\circ}$, is in strong contrast to the observed
$A^1$ dependence ($R$(Au/C$)\approx16$) of the total cross sections at
the same beam energy \cite{pnpi}. The authors of Ref.~\cite{pnpi}
interpreted the latter as an indication for multi-step processes.  A
qualitative explanation \cite{syst} of this discrepancy can be
obtained if one assumes that in the subthreshold domain the large
angle $K^+$-production is due to reactions of the projectile protons
with more than a single nucleon.  The $A$ dependence of the ANKE data
at small angles indicates direct production of $K^+$-mesons through
the interaction of incident protons with surface nucleons.  We have to
conclude, however, that the $A$ dependence of the differential spectra
does not reveal the expected change from direct to multi-step kaon
production with decreasing beam energy.

According to model calculations, the momentum distributions of protons
and deute\-rons, measured in coincidence with the $K^+$-mesons, are
sensitive to the different production mechanisms
\cite{sibirtsev,cassing94}.  A ($K^+d$)-pair in the final state can, for
example, be observed in case of a particular two-step reaction with
deuteron formation in the first step ($pN_1\rightarrow d\pi$, $\pi
N_2\rightarrow K^+\Lambda$) \cite{pnpi,sibirtsev}, while $K^+p$ pairs may
come both from the two-step and direct production.  On the other hand,
the influence of deuteron break-up in the nucleus might be strong,
so that one does not observe any coincident deuterons. 

\begin{figure}[ht]
  \begin{center}
    \resizebox{\figurewidth}{!}{\includegraphics[scale=1]{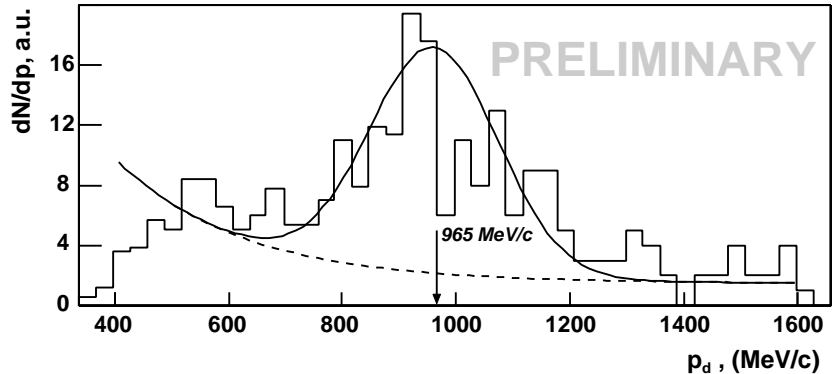}}
    \caption{Coincident deuteron momentum spectra at $T_p=1.2$~GeV.
      The deuteron emission angles were restricted to
      $\vartheta_d<8^\circ$.  The solid line shows the result of a fit
      to the data, the arrow the fitted peak position.}
  \label{fig:K+d}
  \end{center}
\end{figure}

First test measurements of $K^+p$ and $K^+d$ coincidences have been
performed at $T_p=1.2,\,1.5$ and 2.3~GeV. The amount of detected
deuterons is rather large (comparable to the $K^+p$ yield) at
$T_p=1.2$ GeV and decreases towards larger energies.  This is in line
with the interpretation that the contribution of two-step mechanisms
is large at 1.2 GeV and almost negligible at the highest energy.  The
momentum spectrum of the correlated deuterons at $T_p=1.2$ GeV is
shown in Fig.~\ref{fig:K+d}.

A peak structure in the deuteron-momentum distribution is observed at
$p_d\approx 965$~MeV/c, which is close to the expected value for
on-shell two-body kinematics of the reaction $pN\rightarrow d\pi$ with
backward emission of the deuteron in the ($d\pi$) CMS. In this
collinear geometry, pions with maximum momenta are produced which is
energetically favorable for $K^+$-production in the second step. A
similar peak structure has been predicted \cite{sibirtsev} for the
deuteron spectrum at $T_p=1.0$ GeV and, thus, this spectrum can be
considered as first direct experimental evidence of the two-step
production below threshold. However, detailed model calculations still
have to show whether the observed peak structure cannot be faked by
competing processes like coalescence \cite{fastd} or pick-up
reactions. Another source of $K^+d$ pairs might be the kaon production
on two- (or multi-) nucleon clusters \cite{pnpi,roc} which is expected
to yield deuterons with similar momentum distributions as from the
two-step mechanism.

\section{Final-state interaction effects}
\label{sec:fsi}
\begin{figure}[ht]
  \begin{center}
    \resizebox{\figurewidth}{!}{\includegraphics[scale=1]{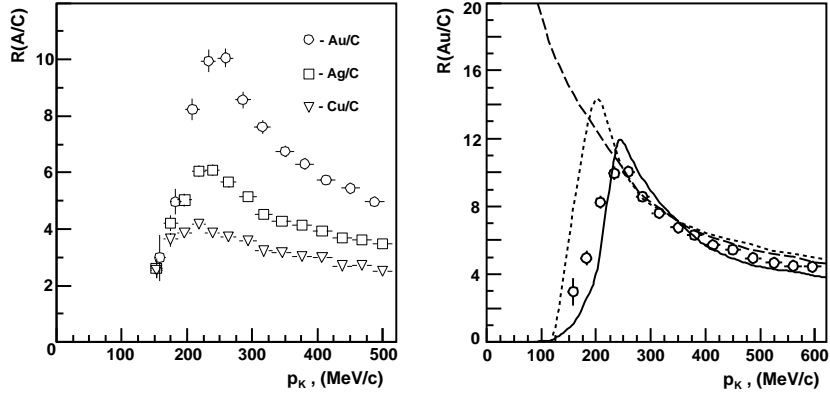}}
    \vspace*{-2mm}
    \caption{Left figure: Ratios of the $K^+$ production cross 
      sections Cu/C, Ag/C, and Au/C measured at $T_p=2.3$ GeV as a
      function of the kaon momentum. Right figure: Ratio for Au/C at
      $T_p=2.3$ GeV.  The dotted line is obtained from transport
      calculations \cite{cassing99,rudy} including only the Coulomb
      potential, the solid line corresponds to calculations with the
      addition of a repulsive kaon potential of 20~MeV (as well as
      baryon potentials).  The dashed line corresponds to simulations
      without Coulomb and nuclear kaon potentials. In all cases $K^+$
      rescattering in the nucleus has been taken into account. The
      figures have been taken from Ref.~\cite{PLB}.}
  \label{fig:fsi}
  \end{center}
\end{figure}

The ratios of kaon-production cross sections for Cu/C, Ag/C and Au/C
measured at 2.3 GeV are presented in Fig.~\ref{fig:fsi}.  All data
exhibit similar shapes, rising steadily with decreasing kaon momenta,
passing a maximum and falling steeply at low momenta.  The position of
the maximum varies with the nucleus, a fit to the data gives
$p_\mathrm{max}(A/\mathrm{C})=245\pm 5$, $232\pm 6$, and $211\pm
6$ MeV/c for Au, Ag, and Cu, respectively.

The same effect at low $p_K$ has been observed for beam energies down
to 1.5 GeV, suggesting that the phenomenon is independent of the
$K^+$-production mechanism and, thus, due to final-state interaction
of the $K^+$ with the residual nucleus.  The peak structure seems not
to be visible in the ratio Au/C at $T_p=1.0$ GeV shown on the l.h.s.\ 
of Fig.~\ref{fig:ratios}. However, at this low beam energy the large
background permitted us to extract the cross sections only down to
$p_K\approx220$ MeV/c, which is slightly below the peak positions
observerd at $T_p\geq 1.5$ GeV.  Taking into account the large error
bar of the lowest data point, the apparent constant ratio at $T_p=1.0$
GeV is not in contradiction with the strong decrease abserved at
higher energies.  

Fig.~\ref{fig:fsi} shows that the position of the maximum in $R(A$/C)
increases with $A$. The situation has a parallel in the well-known
suppression of $\beta^+$ emission in heavy nuclei at low positron
momenta, arising from the repulsive Coulomb potential $V_C$.
Positively charged particles, like $K^+$-mesons, produced at rest at
some radius $R$ in the target nucleus, acquire a momentum
$p=\sqrt{2mV_C(R)}$, if other interactions are negligible.

Moreover, it is known from $K^+$ elastic scattering experiments at
higher energies~\cite{Marlow} that the nuclear $K^+A$ potential is
mildly repulsive, in agreement with one-body optical potentials based
upon low-energy $K^+N$ scattering parameters~\cite{Martin}. At normal
nuclear density, $\rho_0 \approx 0.16\,$fm$^{-3}$, the predicted
repulsive $K^+A$ potential of strength $V_K\approx 20$
MeV~\cite{sibirtsev98} would shift the kaon momenta to higher values.

Results of calculations in the framework of a coupled channel
transport model~\cite{cassing99,rudy} for $R$(Au/C) are shown in
Fig.~\ref{fig:fsi} (r.h.s.).  Without including the Coulomb and
nuclear kaon potentials (dashed line) the ratio exhibits a smooth
momentum dependence with a steady increase towards low momenta as a
result of the stronger $K^+$ rescattering processes for the Au
target.  Coulomb interaction leads to a distortion of the momentum
spectrum and provides a maximum at $p_K\approx 200$ MeV/c (dotted
line).  An additional change is observed in the calculations when a
repulsive kaon nuclear potential is also considered. Using a kaon
potential of $V_K= 20$ MeV in the calculations, a reasonable agreement
with the data is achieved, with a maximum close to the experimental
value of $245$ MeV/c (solid line).  We expect that after refined model
calculations the nuclear $K^+$-potential at normal nuclear density
$\rho_0$ can be determined with an accuracy of better than $\pm3$~MeV.
This is significantly better than the current knowledge from heavy-ion
collisions ($\rho>\rho_0$) where the uncertainties are of the order of
15 MeV \cite{sibirtsev98}.

\section{Outlook}
\label{sec:outlook}
The ANKE experimental program on inclusive production  of
$K^+$-mesons in $pA$ reactions has been finalized. First test
measurements of $K^+p$ and $K^+d$ coincidences have been performed in
2001 and show the predicted \cite{sibirtsev,cassing94} sensitivity to
different reaction mechanisms leading to kaon production below and
above threshold, see discussion in Sect.~\ref{sec:mechanisms}. It is
planned to continue these studies during the year 2003. Similar
measurements have already been made with a hydrogen target
at $T=2.65$ and 2.83 GeV. First analyses of these data showed that
ANKE allows to study the production of heavy hyperons, up to the
$\Lambda$(1520), in $pp\rightarrow pK^+ Y$ reactions
\cite{inpc,gatchina}.

In spring 2002 a detection system for negatively charged particles, in
particular $K^-$-mesons, has been commissioned at ANKE. It will be
used to study the production of $K^+K^-$ pairs in $pp$
\cite{proposal104}, $pn$ \cite{proposal97} and $pA$
\cite{proposal21} collisions. From the latter measurements one hopes to
obtain information about the production and interaction of
$K^-$-mesons in the nuclear medium. The nuclear potential of the
$K^-$-mesons is supposed to be strongly attractive, in contrast to
$K^+$-mesons where we found a repulsive potential, or mass shift, of
$V_K\approx 20$ MeV.  The measurements can be carried out in the end
of 2003 or beginning of 2004.

\section*{Acknowledgments}
This work profitted significantly from discussions with members of the
ANKE collaboration, in particular W.~Cassing, V.~Koptev, Z.~Rudy,
A.~Si\-bir\-tsev and C. Wilkin.  We would like to thank B.L.~Ioffe and
H.~Str\"oher for carefully reading the manuscript.  Support
from the following funding agencies was of indispensable help for
building ANKE, its detectors and DAQ: Georgia (Department of Science
and Technology), Germany (BMBF: grants WTZ-RUS-649-96, WTZ-RUS-666-97,
WTZ-RUS-685-99, WTZ-POL-007-99, WTZ-POL-015-01, WTZ-POL-041-01; DFG:
436 RUS 113/337, 436 RUS 113/444, 436 RUS 113/561, State of
North-Rhine Westfalia), Poland (Polish State Committee for Scientific
Research: 2 P03B 101 19), Russia (Russian Ministry of Science:
FNP-125.03, Russian Academy of Science: 99-02-04034, 99-02-18179a) and
European Community (INTAS-98-500).

\end{document}